\documentclass[aps,prl,reprint,longbibliography]{revtex4-1}
\usepackage{CJK}
\usepackage{graphicx}
\usepackage{hyperref}

\begin{document}
\begin{CJK*}{UTF8}{gbsn}

\title{Spin-flip noise due to nonequilibrium spin accumulation}

\author{Liang LIU (刘亮)}
\author{Jiasen NIU (牛佳森)}
\author{ Huiqiang GUO (郭会强)}
\author{ Jian WEI (危健)}\email{weijian6791@pku.edu.cn}
\affiliation{International Center for Quantum Materials, School of Physics, Peking University, Beijing 100871, China}
\affiliation{Collaborative Innovation Center of Quantum Matter, Beijing, China}
\author{D. L. Li}
\author{J. F. Feng}\email{jiafengfeng@iphy.ac.cn}
\author{X. F. Han}
\affiliation{Beijing National Laboratory of Condensed Matter Physics, Institute of Physics, Chinese Academy of Sciences, Beijing 100190, China}
\author{J. M. D. Coey}
\affiliation{CRANN and School of Physics, Trinity College, Dublin 2, Ireland}
\author{X.-G. Zhang}\email{xgz@ufl.edu}
\affiliation{Department of Physics and the Quantum Theory Project, University of Florida, Gainesville 32611, USA}

\date{\today}

\begin{abstract}
When current flows through a magnetic tunnel junction (MTJ), there is spin accumulation at the electrode-barrier interfaces if the magnetic moments of the two ferromagnetic electrodes are not aligned. Here we report that such nonequilibrium spin accumulation generates its own characteristic low frequency noise (LFN). Past work viewed the LFN in MTJs as an equilibrium effect arising from resistance fluctuations ($S_R$) which a passively applied current ($I$) converts to measurable voltage fluctuations ($S_{V}=I^{2}S_{R}$). We treat the LFN associated with spin accumulation as a nonequilibrium effect, and find that the noise power can be fitted in terms of the spin-polarized current by $S_{I}f=aI\coth(\frac{I}{b})-ab$, resembling the form of the shot noise for a tunnel junction, but with current now taking the role of the bias voltage, and spin-flip probability taking the role of tunneling probability. 

\end{abstract}
\maketitle
\end{CJK*}

Low frequency noise (LFN), often appearing as $1/f$ noise, is known to exist in both AlO$_{x}$-based~\cite{Ingvarsson2000prl,Jiang2004prb} and  MgO-based MTJs~\cite{Stearrett2012prb,Arakawa2012prb}. So far it was believed to be an \textit{equilibrium} noise such as that
observed in various semiconductor devices and disordered metal films~\cite{Dutta1981rmp,Weissman1988rmp,Fleetwood2015ieee}. With the assumption of equilibrium conductance or resistance fluctuations, mobility and carrier number fluctuations are the two apparent reasons while the microscopic origin varies in different cases. Although recently there was evidence~\cite{Arakawa2015prl} of excess shot noise induced by \textit{nonequilibrium} spin accumulation which is proportional to the spin current, this observation seems irrelevant to the general perception that the LFN in magnetic materials is an essentially equilibrium phenomenon, which has no explicit connection to the spin-polarized current. Here we show this general perception is wrong and the LFN in MTJs is indeed driven by the spin-polarized current. 

It was through LFN measurements that Hardner \textit{et al.}~\cite{Hardner1993prb} first demonstrated the fluctuation-dissipation relation (FDR) in a magnetic system, metallic giant magnetoresistance (GMR) multilayers where the LFN peaks at magnetic fields maximizing the GMR derivative. The connection of the FDR to the LFN was later extended  to MTJs~\cite{Ingvarsson2000prl}  by  assuming that magnetic fluctuations in the FM electrodes cause the fluctuations of resistance~\footnote{Note that there is no simple analytical dependence of R(M) as in the GMR case, although in previous works a linear R(M) is often assumed to apply FDR.}. For this reason such field-sensitive LFN is also called magnetoresistive noise, sometimes simply called \textit{magnetic} noise, which can be enhanced by annealing~\cite{Stearrett2010apl}. There is also a field-insensitive LFN ascribed to defect motion or charge trapping in the barrier and/or at the interface between the barrier and electrodes, sometimes called \textit{electronic} noise, which decreases after annealing~\cite{Stearrett2010jap}. This electronic noise is also called barrier resistance noise, which is again an equilibrium effect.

However, we find that MTJs driven far away from equilibrium can have a characteristic LFN distinct from that described by the FDR. Instead, in the presence of nonequilibrium spin accumulation, the noise power spectral density (PSD) of the LFN exhibits shot noise like dependence on the total current~\cite{Blanter2000pr},
\begin{equation}
S_{I}f^{\gamma}=aI\coth\left(\frac{I}{b}\right)-ab,
\label{Eq_fitting}
\end{equation}
where $S_{I}$ is the current noise PSD, $a$ and $b$ are fitting parameters that depend on temperature and magnetic field, and the power law exponent $\gamma$ is close to 1 (and will be considered to be 1 unless otherwise mentioned). This dependence on $I$ cannot be trivially converted to a dependence on bias voltage $V$ because the I-V characteristic (IVC) is highly nonlinear.
In contrast, when the $1/f$ noise in MTJs is treated as an equilibrium resistive noise, its PSD can be described as~\cite{Dutta1981rmp,Weissman1988rmp,Fleetwood2015ieee}
\begin{equation}
S_{R}/R^{2}=S_{V}/V^{2}=\alpha/Af^{\gamma},
\label{Eq_S_V}
\end{equation}
where $A$ is the junction area (sometimes the volume of the FM electrode is used depending on whether the source of the noise is at the interface~\cite{Arakawa2012prb} or inside the electrode~\cite{Stearrett2012prb,Jiang2004prb}), $\alpha$ is an empirical parameter~\cite{Hooge1981rpp}, and again the power law exponent $\gamma$ is close to 1. In conventional LFN theory, ohmic devices with linear I-V characteristic (IVC) are considered and the applied current ($I$) is merely used to convert the resistance fluctuations ($S_{R}$) to measurable voltage fluctuations, i.e., $S_{V}=V^{2}S_{R}/R^{2}=I^{2}S_{R}$. In other words, current itself does not introduce any noise beyond serving as a probe for the equilibrium resistance noise, in contrast to Eq.~(\ref{Eq_fitting}) where current explicitly affects LFN owing to nonequilibrium spin accumulation. 

\begin{figure}
\includegraphics[width=9cm]{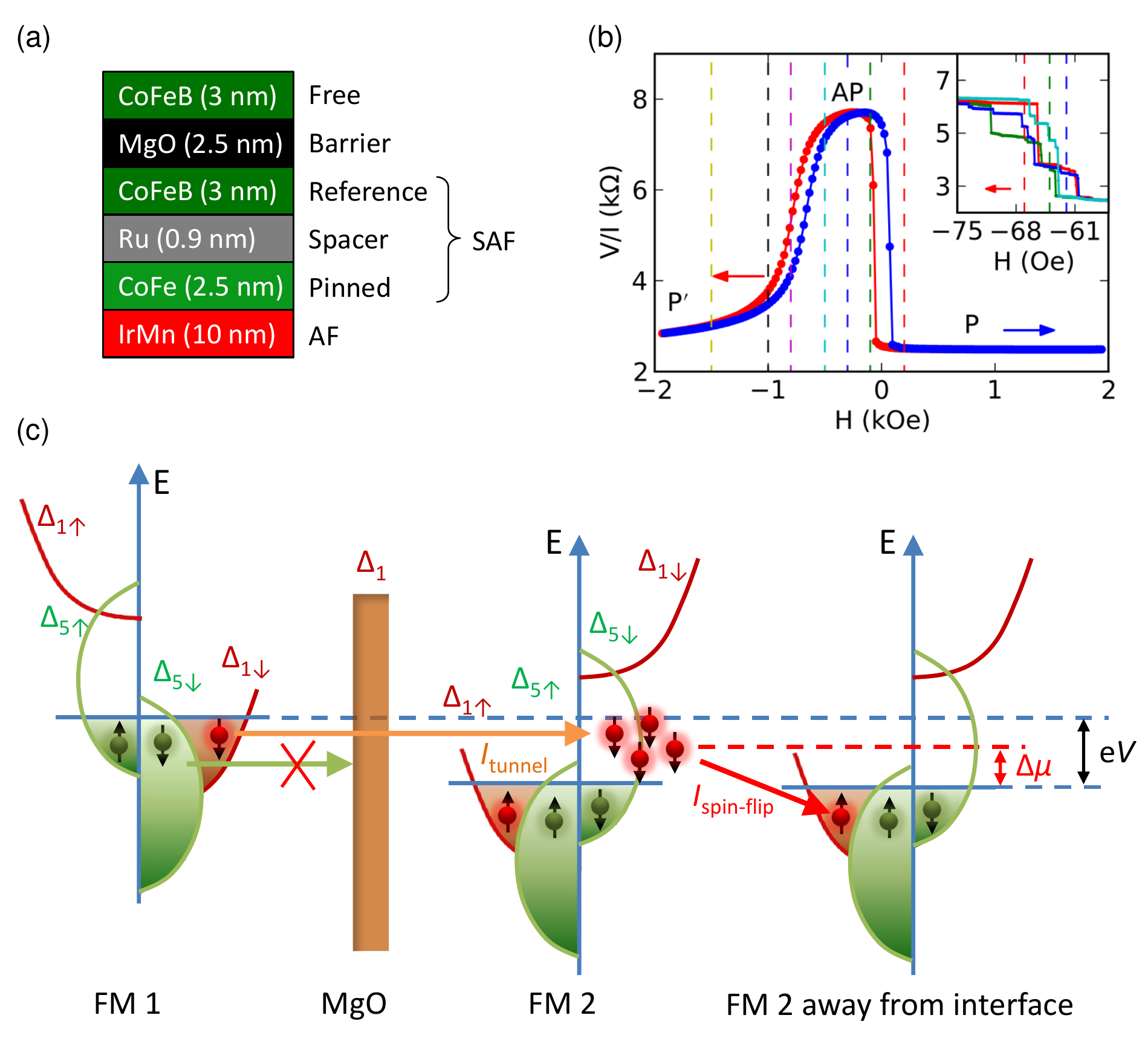}
\caption  {\small  (color online) (a) A simple schematic of the MTJ stack. (b) Magnetoresistance (MR) with field ramping up (blue) and down (red) at a fixed bias current of 6.5 $\mu$A at 296 K for sample M8. The vertical dashed lines denote different fields (ramping down) where noise measurements are shown in Fig.~\ref{Fig_bias_dependence_diff_field}. P, AP and P$^{\prime}$ denote parallel, antiparallel, and another parallel state respectively. The inset shows four continuous runs near the free layer reversal field (ramping down) for sample M9. Three fields (-62 Oe, -64 Oe and -67 Oe) are chosen for noise measurements shown in Fig.~\ref{Fig_bias_dependence_free_layer}. (c) Schematic of the spin accumulation in antiparallel state. With the assumption of ideal symmetry filtering of  the MgO barrier, the down-spin  $\Delta_{1}$ electrons from the left side dominate the tunnel process at large bias. In the right side there is no available down-spin $\Delta_{1}$ states, so these electrons undergo spin-flip scattering and become the up-spin $\Delta_{1}$ states. The \textit{nonequilibrium} spin accumulation is characterized by chemical potential difference $\Delta\mu$ between up and down-spin $\Delta_{1}$ states. The DOS energy diagram follows that of the calculated band dispersions~\cite{Bagayoko1983prb,Yuasa2007jpd} but not to the scale. }
\label{Fig_model}
\end{figure}

The nonequilibrium spin accumulation is illustrated schematically in Fig.~\ref{Fig_model}c, and Fig.~\ref{Fig_model}a shows a simple schematic of a MTJ stack with the free layer, barrier, reference layer and exchange bias layers. With ideal symmetry filtering of the MgO barrier at large bias, majority $\Delta_{1}$ electrons dominate the tunnel process~\cite{Butler2001prb,Zhang2004prb}, while $\Delta_{5}$ electrons are blocked by the MgO barrier. In the high-resistance antiparallel (AP) state (indicated in Fig.~\ref{Fig_model}b), since there are no available minority $\Delta_{1}$ states on the right side, to have a finite current the $\Delta_{1}$ electrons from left side need to be spin-flipped into the majority $\Delta_{1}$ states on the right side. This spin-flip process generates current ($I_{spin-flip}=I_{tunnel}$ in the ideal case) as well as current noise, and could contribute to the reduction of tunnel magnetoresistance ratio with increasing bias as more spin-flip processes can be activated. A finite chemical potential difference $\Delta\mu$ between up and down-spin $\Delta_{1}$ states can be assumed to describe the nonequilibrium spin accumulation at the interface, which is current-dependent instead of voltage-dependent~\footnote{See the Supplementary Material for the derivation of current-dependent noise,  discussion of the magnetic after effect, the comparison of the $S_{V}$ and $S_{I}$ in the P state, the change of $\gamma$ at 3.7 K.}.

\begin{figure}
\includegraphics[width=9cm]{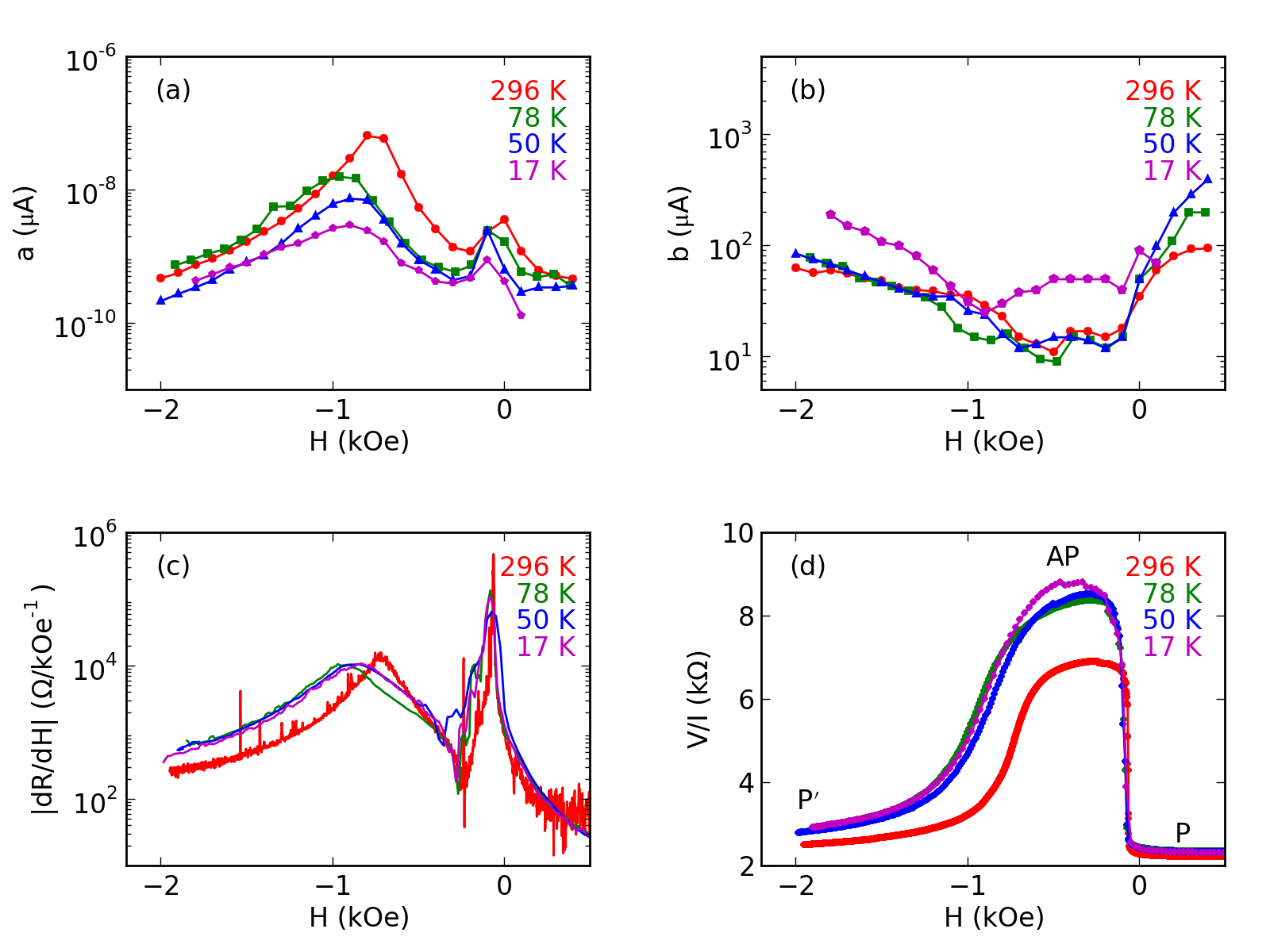}
\caption  {\small  (color online) Field dependence of the fitting parameters $a$ (a) and $b$ (b) following Eq.~\ref{Eq_fitting} at different temperatures. Note that $H_{\Lambda}$, the position of the broad peak, moves when temperature decreases from 296 K to 78 K. The MR measured with small bias (d) and field derivative of MR (c) at different temperatures.  }
\label{Fig_fitting_parameters}
\end{figure}

This unusual result of spin-flip noise following Eq.(\ref{Eq_fitting}) also sheds new light on the unsatisfactory FDR interpretation of previous magnetic noise measurements~\cite{Ozbay2009apl,Stearrett2012prb,Arakawa2012prb,Feng2012jap_a}. Initially the quasi-equilibrium assumption was introduced  for MTJs with low magnetoresistance (MR) and small nonlinearity of the IVC~\cite{Ingvarsson2000prl},  based solely on the observation of $1/f$ noise, as $1/f$ noise is usually found in equilibrium systems. However, the quadratic bias dependence for equilibrium resistive noise suggested by Eq.~(\ref{Eq_S_V}),  $S_{V}\propto V^{2}$, was rarely verified except for the electronic noise at low bias~\cite{Nowak1999apl,Ren2004prb}. Nevertheless, $\alpha$ estimated at some arbitrarily chosen bias was used to characterize the magnetoresistive noise. Then following FDR one expects a linear relation between $\alpha$ and the derivative of MR ($(1/R)dR/dH$)~\cite{Ingvarsson2000prl,Ren2004prb}, or the magnetoresistance-sensitivity product MSP ($\equiv(\Delta R/R^2)(dR/dH)$)~\cite{Ozbay2009apl,Stearrett2012prb,Arakawa2012prb,Feng2012jap_a}. This expected linear relation exists only within a limited field range, which is unsatisfactory and puzzling. Compared to $\alpha$, Eq.(\ref{Eq_fitting}) gives a more accurate description of the noise power over the entire bias range, and the fitting parameter $a$ (see Fig.~\ref{Fig_fitting_parameters}) may replace $\alpha$ to some extent in the FDR.

\begin{figure}
\includegraphics[width=9cm]{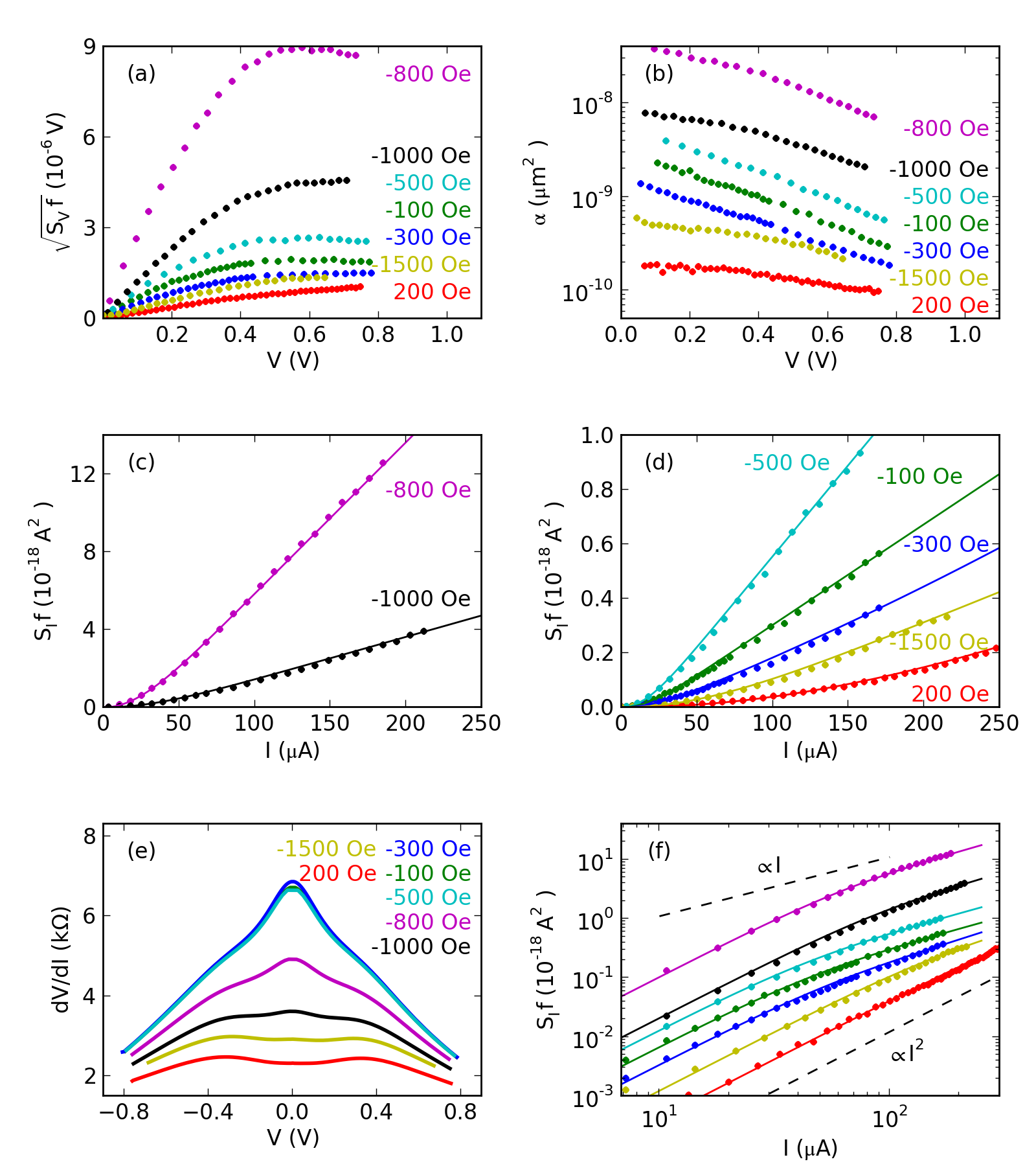}
\caption  {\small  (color online) Bias dependence of normalized power spectrum density  $\sqrt{S_{V}f}$ (a) and $\alpha=AS_{V}f/V^{2}$  (b) at different magnetic field points and at 296 K. The current noise $S_{I}f$  in (c) and (d) are derived by  $\sqrt{S_{V}f}$ in (a)  and  $dV/dI$ in (e), which shows an almost linear dependence on the bias current without saturation. The y scale in (c) is larger than that in (d). In (e), nonlinearity around 0.4 V can be seen for $dV/dI$. (f) Bias dependence of $S_{I}f$ in double-log scale. The slope approaches one at high bias except for the P (200 Oe) and P$^{\prime}$ states (-1500 Oe). The solid lines are fits following Eq.~\ref{Eq_fitting}}
\label{Fig_bias_dependence_diff_field}
\end{figure}

The magnetic noise should be understood in the context of the magnetoresistance, which for our CoFeB/MgO/CoFeB MTJ sample is shown in Fig.~\ref{Fig_model}b. Measured with a small excitation current, the resistance changes more than 200\% from the P to AP states (P, AP denote parallel, antiparallel states respectively). The reference layer is pinned by the synthetic antiferromagnetic (SAF) layer, and its hysteresis loop is exchange biased to around -800 Oe, above which the magnetization of the reference layer also aligns with the applied field and another parallel state P$^{\prime}$ is resulted. The hysteresis loop near zero field is due to the free layer. Details related to sample fabrication and noise measurements can be found in Ref~[\onlinecite{Liu2014prb,Liu2014aipa}], and data presented here are for similar devices M8 and M9 on the same substrate. 

Previous noise measurements and FDR analyses focused on the range of the magnetic field where reference layer reversal occurs (from AP to P$^{\prime}$ as shown in Fig.S.1 and Fig.~\ref{Fig_fitting_parameters}d). The derivative of the MR within this range is not as large as near the free layer reversal regime, allowing a quasi-equilibrium state to be assumed and then the conventional Eq.~(\ref{Eq_S_V}) was applied to find $\alpha$. The results of such conventional analyses for our samples are shown in Figs.~\ref{Fig_bias_dependence_diff_field}a and \ref{Fig_bias_dependence_diff_field}b.  We see first increase and then saturation of $\sqrt{S_{V}f}$ above roughly 0.4 V in Fig.~\ref{Fig_bias_dependence_diff_field}a.  Indeed, the extracted parameter $\alpha=AS_{V}f/V^{2}$ is not a constant as required by the conventional theory, but shows a strong dependence on the voltage as seen in Fig.~\ref{Fig_bias_dependence_diff_field}b (note the semi-log scale), indicating deviation from the resistive noise picture. Similar suppression of $\alpha$ with bias was observed in previous works but no modification of the conventional  analyses was made.~\cite{Gokce2006jap,Almeida2008ieee,Feng2012jap_a,Aliev2014prl}  Even for the P (200 Oe) and P$^{\prime}$ (-1500 Oe) states, where electronic noise is supposed to dominate, we can still see strong deviation from a constant $\alpha$ in Fig.~\ref{Fig_bias_dependence_diff_field}b.

Rather than relying on the differences between the magnetic and electronic noises to explain the data, we find that Eq.~(\ref{Eq_fitting}) fits equally well for all cases at different fields without the need to make a distinction between different types, as shown in Figs.~\ref{Fig_bias_dependence_diff_field}c and \ref{Fig_bias_dependence_diff_field}d (note the different $y$ scales).
Here we have taken into account quantitatively the nonlinear IVC (as demonstrated by $dV/dI$ in Fig.~\ref{Fig_bias_dependence_diff_field}e) and  plot the current noise $S_{I}f=S_{V}f/(dV/dI)^{2}$. The striking feature is that there is no longer saturation, suggesting that $S_{I}f$, rather than $S_{V}f$, represents the intrinsic noise. This current noise increases linearly with $I$ above a threshold value, corresponding to the fitting parameter $b$ for each case, e.g., about 25 $\mu$A  for -800 Oe.  In Fig.~\ref{Fig_bias_dependence_diff_field}f,  $S_{I}f$ versus $I$ is replotted in double-log scale, such that all data points can be presented and the power exponent of $I$ can be identified. We can see at high bias the slope approaches 1, except for the 200 Oe (P) and -1500 Oe (P$^{\prime}$) data, for which the noise is smaller and the slope is close to 2. These two limiting cases can be described well by Eq.~(\ref{Eq_fitting}) since for $I/b \ll 1$, $\coth(I/b)\approx (I/b)^{-1}+(I/b)/3$, then $S_{I}f \approx aI^{2}/3b$;  and for $I/b \gg 1$, $\coth(I/b)\approx 1$, then $S_{I}f \approx a(I-b)$.  In fact, even for the 200 Oe (P) data, $S_{I}f \propto I^{2}$ fits better than the conventional $S_{V}f \propto  V^{2}$ (See Fig.S1 in the Supplementary Material).

The similarity between the observed $S_{I}f$ scaling and the shot noise formula for a tunnel junction is striking. The latter is of the form~\cite{Blanter2000pr,Arakawa2011apl,Arakawa2012prb}
\begin{equation}
S_{I}=\frac{4k_{B}T}{dV/dI}+2F\left[eI\coth\left(\frac{eV}{2k_{B}T}\right)-\frac{2k_{B}T}{dV/dI}\right],
\label{Eq_shot_noise}
\end{equation}
where $F$ is the Fano factor , $k_{B}$ is the Boltzmann constant. However, there are fundamental differences between the observed $S_{I}$ and electronic shot noise, the most important ones being that the noise power in Eq. (\ref{Eq_shot_noise}) is independent of the frequency and its scaling with the voltage applied on the junction. In sharp contrast, in Eq. (\ref{Eq_fitting}) the noise power scales with $1/f$ and depends only on the current, excluding a possible tunneling origin. There is no thermal noise term in Eq. (\ref{Eq_fitting}) while it is the first term in Eq.~(\ref{Eq_shot_noise}) and is always there even if the shot noise is absent (when $F=0$, the second term disappears). A simple model is described in the Supplementary Material where Eq.~(\ref{Eq_fitting}) is derived from the consideration of spin accumulation and a spin-flip current in the AP state, as already illustrated in Fig.~\ref{Fig_model}c. This model differs from past theoretical proposals that via spin-transfer torque, spin-current noise can cause enhanced magnetization fluctuations at high current density which are reflected by resistance noise.~\cite{Foros2005prl,Foros2009prb} Here we assume a continuous distribution of spin-flip scattering rates and individual particle nature(angular moment is quantized) of the spin-flip events  that lead to both the 1/f dependence and the shot noise like features.

\begin{figure}
\includegraphics[width=9cm]{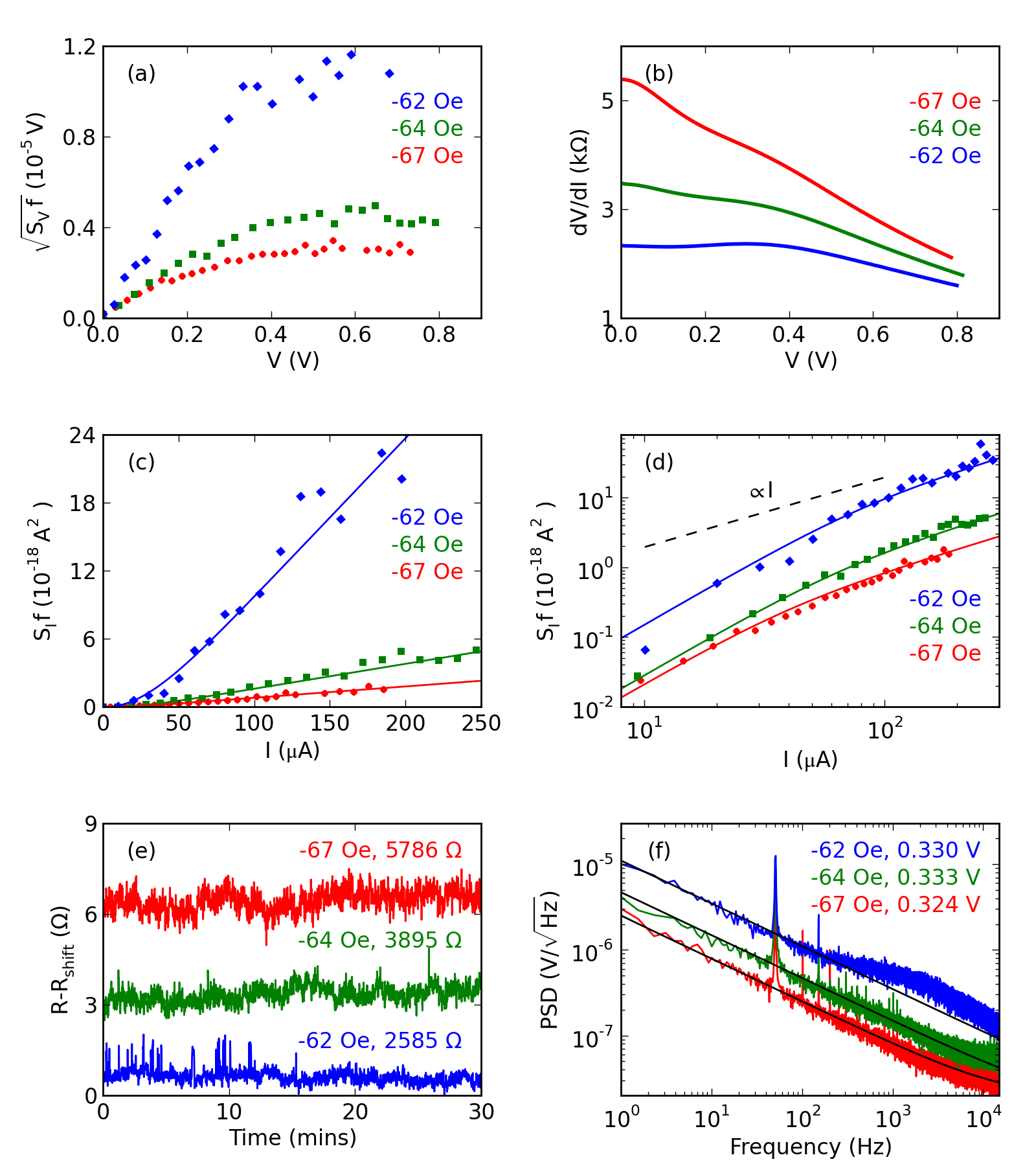}
\caption  {\small  (color online) Bias dependence of normalized power spectrum density   $\sqrt{S_{V}f}$ (a) and $dV/dI$   (b) for three magnetic field points: -62 Oe (blue), -64 Oe (green) and -67 Oe (red) near free layer reversal field at 296 K. Some scattering of the data is caused by small resistance jumps and a few outliers are removed. In (c) the current noise shows a linear dependence on bias current, which is better shown in (d) with double-log scale. The solid lines are fits following Eq.~\ref{Eq_fitting}. (e) Time traces of DC resistance (shifted for clarity) at three different magnetic fields in different runs, demonstrating the stability of the resistance state. (f) Typical noise spectra near 0.3 V bias voltage. The black lines are fitting considering both $1/f$ and white noises (electronic shot noise and thermal noise). }
\label{Fig_bias_dependence_free_layer}
\end{figure}

We observe a similar bias dependence of $S_{I}f$ near the free layer reversal regime, where derivative of MR is large and stationary noise measurement was previously considered very difficult\cite{Stearrett2010apl}. Indeed, as shown in the inset of Fig.~\ref{Fig_model}b, there are discontinuous resistance jumps likely due to domain wall jumps instead of reversible domain rotations. However, we find that when the field is stabilized with an electromagnet, the resistance can be stable for a sufficiently long time for noise measurements, e.g., in Fig.~\ref{Fig_bias_dependence_free_layer}e the resistance drift is less than 1 $\Omega$ in a 30-minute time period. In Fig.~\ref{Fig_bias_dependence_free_layer}a, saturation of  $\sqrt{S_{V}f}$ similar to that in Fig.~\ref{Fig_bias_dependence_diff_field}a is shown, albeit with more scatter in the data. The linear dependence of $S_{I}f$ on $I$ is again confirmed in Fig.~\ref{Fig_bias_dependence_free_layer}c (linear scale) and Fig.~\ref{Fig_bias_dependence_free_layer}b (double log scale) by taking into account $dV/dI$ (Fig.~\ref{Fig_bias_dependence_free_layer}d). As shown in  Fig.~\ref{Fig_bias_dependence_free_layer}f,  when resistance jumps are rare, the noise spectra can be well described by $1/f$ line shape (the magnetic after effect is not important, as discussed in the Supplementary Material).

\begin{figure}
\includegraphics[width=9cm]{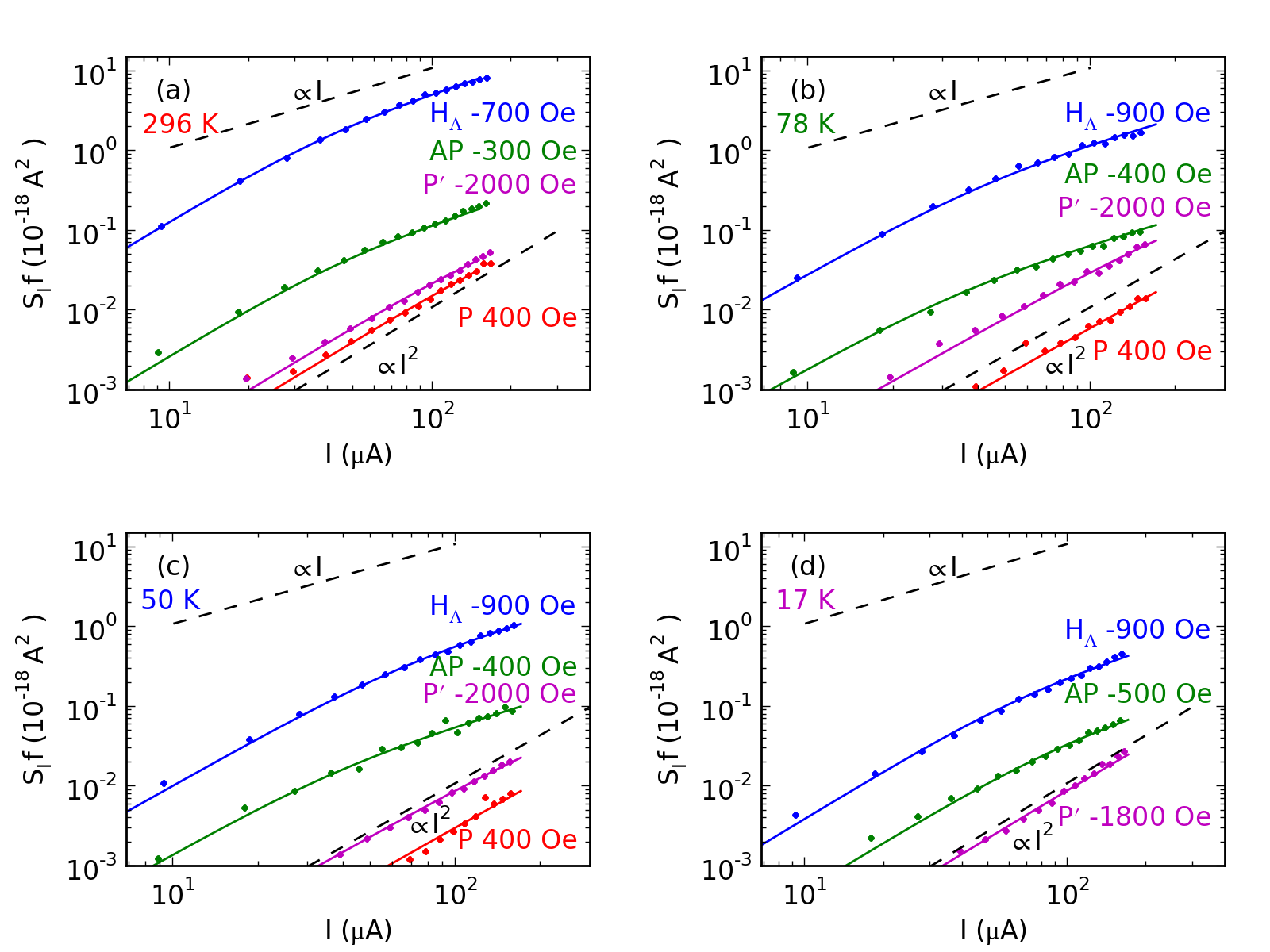}
\caption  {\small  (color online) Bias dependence of normalized power spectrum density   $S_{I}f$ at different temperatures: (a) 296 K;  (b) 78 K; (c) 50 K; (d) 17 K. The solid lines are fits following Eq.~\ref{Eq_fitting}. The 400 Oe data at 17 K are not shown because the $1/f$ noise is too small to be reliably extracted from the background noise. }
\label{Fig_temperature_dependence}
\end{figure}

If FDR were strictly followed, the PSD of equilibrium noise should scale linearly with the thermal fluctuation energy  $k_{B}T$~\cite{Ozbay2009apl,Stearrett2012prb,Arakawa2012prb,Feng2012jap_a}. This linear dependence of $\alpha$ on temperature was not observed, instead, thermal activation at high temperature and saturation  at low temperature were observed for P and AP states~\cite{Jiang2004prb,Gokce2006jap}, and the saturation is suspected to be due to quantum tunneling~\cite{Jiang2004prb}. In general~\cite{Dutta1981rmp,Weissman1988rmp,Fleetwood2015ieee}, LFN is not expected to scale linearly with temperature because the usual microscopic origins of the $1/f$ noise, such as activation of scattering due to defects or impurities, do not have linear temperature dependence.  

LFN at four different temperatures, 296, 78, 50, and 17 K  are presented for four characteristic magnetic states: $H_{\Lambda}$, P, AP, and P$^{\prime}$ in Fig.~\ref{Fig_temperature_dependence}. $H_{\Lambda}$ is the field where LFN shows a broad peak, which is close to the exchange bias field $H_{ex}$ (also called pinning field) of the reference layer, and changes with temperature.  For clarity, data at other fields are not presented in Fig.~\ref{Fig_temperature_dependence} but all fitted parameters are summarized in Fig.~\ref{Fig_fitting_parameters}. From 296 K to 78 K, for all three fields except that for the P$^{\prime}$ state, one can clearly see the decrease of the noise. The decrease is especially clear for the $H_{\Lambda}$ state, which moves from around -700 Oe to -900 Oe, as can be inferred in Fig.~\ref{Fig_fitting_parameters}a. The typical AP state also moves from -300 Oe to -400 Oe and then to -500 Oe with decreasing temperature (see MR in Fig.~\ref{Fig_fitting_parameters}d). In fact,  the increase of LFN at -2000 Oe (P$^{\prime}$ state) from 296 K to 78 K should be explained by the change of magnetic state.  

With our model, the decrease of LFN with temperature implies that the spin-flip rate  is reduced and the total tunnel current is reduced as well. This is qualitatively consistent with the  reduction of $S_{I}f$ with bias (Figs.~\ref{Fig_temperature_dependence}a and b, as well as Fig.~\ref{Fig_fitting_parameters}a) and with the resistance increase in the AP state (Fig.~\ref{Fig_fitting_parameters}d). Below 78 K, the noise decreases slowly and the the AP state resistance increases slowly as well.  The noise decrease is clearer for the $H_{\Lambda}$ state, where the activation energy for spin-flip is minimized~\cite{Ingvarsson2000prl,Koch2000prl,Zhong2011jap,Endean2014apl}. Below 17 K the $\gamma$ parameter started increasing from 1 (see Fig.S2 in the Supplementary Material), thus $S_{I}f$ is no longer good to characterize LFN and this requires further study.

Clear field dependence of LFN, indicated by the fitting parameters $a$ and $b$ as shown in Fig.~\ref{Fig_fitting_parameters},  excludes any influence of interband scattering~\cite{Aliev2014prl} which is field independent and bias dependent. In addition, observation of full electronic shot noise at higher frequencies excludes any influence by localized states within the barrier which would lead to a reduction of the Fano factor. Compared with the electronic shot noise, fitting parameter $a$ represents the particle-like feature of spin-flip noise, resembling charge $e$  and to some extent plays the role of the empirical parameter $\alpha$. There is indeed some correlation between $a$ and the field derivative of MR ($dR/dH$), as can be seen from the similarity between Figs.~\ref{Fig_fitting_parameters}a and \ref{Fig_fitting_parameters}c, although their temperature dependence is not the same. Note that $a$ is determined within a wide current range while $dR/dH$ is measured at zero bias limit.  The parameter $b$ resembles $k_{B}T$ in the expression of electronic shot noise. It indicates the threshold current above which a linear dependence of $I$ is prominent, i.e., the nonequilibrium noise due to spin accumulation dominates.  In Fig.~\ref{Fig_fitting_parameters}b,  $b$ is higher in P and P$^{\prime}$ states since when $\Delta_{1}$ majority electrons can tunnel to $\Delta_{1}$ majority available states,  there is no spin accumulation and $I_{spin-flip}$ can not dominate $I_{tunnel}$.   There is little temperature dependence for $b$ until 17 K where two small local minima appear near the reversal fields, which may be again explained by that the activation energy barrier for spin-flip process is minimized near the reversal fields, thus $I_{spin-flip}$ is enhanced. 

In conclusion, we find that the $1/f$ noise in MTJs is better described by the current noise $S_{I}f$ with bias dependence similar to shot noise, rather than by the conventional resistance noise ($S_{V}=I^{2}S_{R}$). The origin of this noise is traced to spin accumulation at the interface and the subsequent spin-flip current due to the highly spin-polarized tunneling current. The particle nature of individual spin-flip events and the time-scale distribution of such events combine to generate a spin-flip shot noise that is distinct from the shot noise of a tunnel junction and has the $1/f$ frequency scaling.  Our result may shed new light on spin injection, spin detection, and other spin-dependent devices involving spin accumulation. 

Work at Peking University was supported by National Basic Research Program of China (973 Program) through Grant No. 2011CBA00106 and  2012CB927400, as well as National Natural Science Foundation of China [NSFC, Grant No. 11474008]. Work at IOP, CAS was supported by the State Key Project of Fundamental Research of Ministry of Science and Technology [MOST, No. 2014AA032904], and NSFC, Grant Nos. 11434014 and 51229101. Work at Trinity College  Dublin was supported by SFI Contract 10/IN1.13006. XGZ acknowledges support in part by NSF grant ECCS-1508898. JW would like to acknowledge discussions about magnetic excitations with Fa Wang, Shindou Ryuichi, Wei Han, and also acknowledge Jiang Xiao for critical reading of the manuscript and discussions about the model. 


%

\end{document}